\documentclass[prb,twocolumn,showpacs,preprintnumbers,superscriptaddress,amsmath,amssymb,aps,10pt]{revtex4-1}

\usepackage{units}
\usepackage{graphicx}

\begin{document}

\title{Extension of the Anderson impurity model for finite systems: Band gap control of magnetic moments}

\author{Konstantin~Hirsch}
\email{konstantin.hirsch@helmholtz-berlin.de}

\affiliation{Institut f\"ur Methoden und Instrumentierung der Forschung mit Synchrotronstrahlung, Helmholtz-Zentrum Berlin f\"ur Materialien und Energie GmbH, Albert-Einstein-Stra{\ss}e 15, 12489 Berlin, Germany}

\author{Julian~Tobias~Lau}
\affiliation{Institut f\"ur Methoden und Instrumentierung der Forschung mit Synchrotronstrahlung, Helmholtz-Zentrum Berlin f\"ur Materialien und Energie GmbH, Albert-Einstein-Stra{\ss}e 15, 12489 Berlin, Germany}

\author{Bernd~von~Issendorff}
\affiliation{Physikalisches Institut, Universit\"at Freiburg, Stefan-Meier-Stra{\ss}e 21, 79104 Freiburg, Germany}

\begin{abstract}
We study the spin magnetic moment of a single impurity embedded in a finite-size non-magnetic host exhibiting a band gap. The calculations were performed using a tight-binding model Hamiltonian.
The simple criterion for the magnetic to non-magnetic transition as given in the Anderson impurity model breaks down in these cases. We show how the spin magnetic moment of the impurity that normally would be quenched can be restored upon introducing a gap at the Fermi level in the host density of states. The magnitude of the impurity spin magnetic moment scales monotonically with the size of the band gap. This observation even holds for a host material featuring a strongly discretized density of states. Thus, it should be possible to tune the magnetic moment of doped nano-particles by varying their size and thereby their band gap.
\end{abstract}


\preprint{}

\date{\today}

\maketitle
\section{Introduction}
Materials, especially metals, sparsely doped with magnetic impurities have been widely investigated throughout the last 50 years primarily because some of them display 
the fascinating Kondo effect \cite{Kondo1964}. Such systems can serve as an ideal playground to explore the underlying many body physics. However, the prerequisite to observe a Kondo effect at all is the survival of a local moment of the embedded impurity.\\
The interaction of the localized impurity electronic states with the itinerant electrons of the host material is theoretically described by the well established $sd$- or Anderson model \cite{Anderson1961a}. Within this model the size of the local moment arises from an intricate interplay of the on-site Coulomb repulsion $U_0$, the energy penalty for adding a second electron to the localized state, and the width $2\Gamma$ of the localized state. The width of the localized state, also known as virtual bound state, results from hybridization of the electronic states of the impurity with the delocalized states of the host material. In case of a symmetric arrangement of the impurity spin levels $E_{d,\pm}$, \textit{i.e.} $E_{d,\pm}=E_F\mp U_0/2\left(n_+-n_-\right)$ ($E_F$ is the Fermi energy of the system and $n_\pm$ the occupation of $E_{d,\pm}$), a simple criterion for the existence of a local magnetic moment can be derived \cite{Anderson1961a}:
$U_0/\Gamma>\pi$.\\
In recent years, advances in experimental techniques and theoretical methods allowed to push the research into the finite size domain \cite{Skomski2010,Pastor2005,Kaul2009,Kaul2006,Kaul2005,Rotter2009,Liu2012,Thimm1999,Cornaglia2002,Booth2005}. In finite systems the itinerant electrons are confined and populate highly discretized energy levels. This in turn can be expected to have tremendous influence on the description within the Anderson impurity model, which accounts only for a continuous host density of states. Indeed we found evidence that the size of the spin magnetic moment of a chromium impurity embedded in a small gold cluster is strongly affected by the discretized density of states of the host particle \cite{Hirsch2013}. This becomes most evident in host particles that exhibit a shell closure and therefore a wider highest-occupied--lowest-unoccupied molecular orbital (HOMO-LUMO) gap. 
To get a more fundamental grasp on the influence of an energy gap or a highly discretized host density of states on the spin magnetic moment of an embedded impurity, we investigate such a systems 
using a modified Anderson impurity model.
\section{Model Hamiltonian}
We model the system in a tight-binding approach, using the following model Hamiltonian:
\begin{equation}
\mathcal{H}_{TB}=\begin{pmatrix} E_{d,\pm} & a & \cdots & a \\ a & E_{k,1} & 0 & 0\\  \vdots & 0& \ddots & 0 \\ a& 0& 0&  E_{k,N} \end{pmatrix}
\label{eq:TB_Hamiltonian}
\end{equation}
In $\mathcal{H}_{TB}$, a single localized orbital at energy $E_{d,\pm}$ interacts with a finite number $N$ of delocalized states at energies $E_{k,i}$. Like in the Anderson model, the coupling strength $a$ of the localized orbital to the continuum states is assumed to be the same for all states $E_{k,i}$. Diagonalization of the matrix $\mathcal{H}_{TB}$ yields the $N+1$ eigenstates $\phi_i$ and eigenenergies $\epsilon_i$ of the system. This is to be done separately for majority $(+)$ and minority $(-)$ impurity spin states $E_{d,\pm}= E_F \mp U_0/2\left(n_+-n_-\right)$ to yield spin resolved eigenfunctions $\phi_i^\pm=(c_1^{i,\pm},c_2^{i,\pm}, \ldots c_{N+1}^{i,\pm})$ and eigenenergies $\epsilon_i^\pm$. The states $E_{d,\pm}$ are separated by the on-site Coulomb repulsion $U_0$, which is the energy neccessary to add an electron to the localized orbital. Eigenfunctions and eigenenergies obtained from diagonalization are used to calculate the occupation numbers $n_\pm$ of the majority and minority spin states from the projected spin density of states $\rho_\pm\left(E\right)$ as:
\begin{eqnarray}
\rho_\pm\left(E\right) &=&\sum_i|c_1^{i,\pm}|^2 \, \delta (E-\epsilon_i) \label{eq:MAIM_DOS}\\
n_\pm &=& \int_{- \infty}^{E_F}\rho_\pm\left(E\right)dE \label{eq:MAIM_occ}
\end{eqnarray}
Here, $\delta\left(E\right)$ is the delta function and $E_F$ the Fermi energy of the system.\newline
\begin{figure}[t]
  \includegraphics[width=0.9\columnwidth]{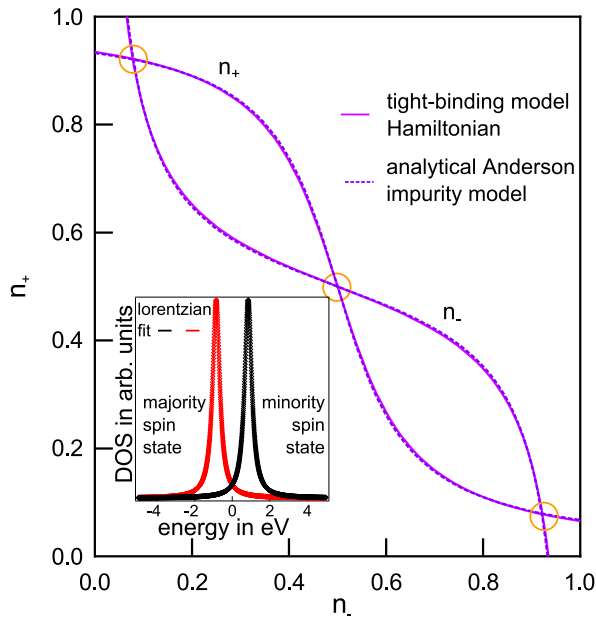}
  \caption{Comparison of the solutions for the occupation numbers $n_\pm$ of spin-up and -down states using $U_0/\Gamma \approx 9.1$, obtained analytically from the Anderson impurity model and the tight-binding Hamiltonian equation (\ref{eq:TB_Hamiltonian}), using a coupling strength $a=\unit[0.02]{eV}$, on-site Coulomb repulsion $U_0=\unit[2]{eV}$ and a host density of states of $\unit[180]{eV^{-1}}$. Both models yield almost identical results, the self-consistent solutions are marked by the orange circles. Inset: Impurity state projected density of states for the spin polarized solution, obtained by solving equations (\ref{eq:TB_Hamiltonian}-\ref{eq:MAIM_occ}) self-consistently. The Lorentzian fit agrees well with the density of states.}
  \label{fig:Vgl_AIM_TB}
\end{figure}
In order to find the spin polarization $\frac{\left(n_+ - n_-\right)}{\left(n_+ + n_-\right)}$ of the system the equations (\ref{eq:TB_Hamiltonian}-\ref{eq:MAIM_occ}) have to be solved self-consistently, since the energetic position of the localized orbital depends on the occupation $n_\pm$ and vice versa. More specifically, the energetic position of $E_\mp$ is determined by $n_\pm$ which in turn dictates the occupation of $n_\mp$. Like Anderson \cite{Anderson1961a} we solve this problem graphically by plotting the majority spin state occupation as a function of the minority spin state occupation $n_+\left(n_-\right)$ as well as $n_-\left(n_+\right)$. A self-consistent solution is found at the intersections of both curves, as shown in Fig. \ref{fig:Vgl_AIM_TB}. We test our model by comparing its results for dense but discrete levels, approximating a continuous band, to the analytical solution of the Anderson impurity model. In this limit both models should yield identical results. We chose a constant density of states of $\unit[180]{eV^{-1}}$, which is comparable to the density of states at the Fermi level of a free electron gas as can be found, for example, in a gold Au$_{660}$ nano-particle if the level bunching due to electron shell effects is neglected. The analytical solution \cite{Anderson1961a} of the Anderson impurity model for the occupations of majority and minority spin state is given by
\begin{equation}
n_\pm=\frac{1}{\pi}\arctan\left(\frac{U_0\cdot \left(n_\mp-0.5\right)}{\Gamma}\right)+0.5.
\label{eq:AIM_occ}
\end{equation} 
A symmetrical arrangement of the impurity states $E_{d,\pm}$ relative to the Fermi energy $E_F$ is assumed. Such a symmetrical arrangement of the levels is a reasonable assumption implying that the dopant remains charge neutral. Although in metallic systems the impurity can be charged to some extent, this will be well below one elementary electric charge, rendering its influence on the Anderson model negligible.\newline
Fig. \ref{fig:Vgl_AIM_TB} demonstrates that the numerical solution of $\mathcal{H}_{TB}$ in the continuous band limit and the analytical solution of the Anderson impurity model are nearly indistinguishable. Furthermore, the inset of Fig. \ref{fig:Vgl_AIM_TB} shows the impurity state projected density of states that results from the numerical calculation. The Lorentzian shape of the virtual bound states is also in very good agreement with what one would expect from the Anderson impurity model and further confirms that our tight-binding model agrees with the Anderson impurity model in the continuous band limit.\newline
The on-site Coulomb repulsion was chosen to be $U_0=\unit[2]{eV}$, since typical values of $U_0$ for transition metals are ranging from \unit[1]{eV} to \unit[6]{eV} \cite{Sasoglu2011,Kulik2006,Anisimov1991a}. For a given density of states and on-site Coulomb repulsion the half-width of the virtual bound state is solely determined by the coupling strength $a$, which was set  to $\unit[0.02]{eV}$ here. This set of parameters results in a width $2\Gamma$ of $\approx \unit[0.44]{eV}$ which is obtained from a lorentzian fit shown in the inset of Fig. \ref{fig:Vgl_AIM_TB} and compares well with the analytical value $2 \Gamma= \pi a^2 \rho(E_F)=\unit[0.45]{eV}$. Generally, a parameter range of the coupling strength $a$ of $\unit[0.02]{eV}-\unit[0.08]{eV}$ yields line widths which are consistent with line widths seen in UPS experiments carried out on $3d$-transition metal impurities embedded in gold and silver \cite{Reehal1980,Hochst1980,Hillebrecht1983,Folkerts1987}, scanning tunneling experiments on adatoms \cite{Crommie1993} as well as density functional theory calculations \cite{Podloucky1980,Weissmann1999}.\newline
It should be noted that the model introduced here is constructed for a single impurity state only. An extension to multi-orbitals as present in, e.g., $3d$-transition metals can be done, but does not fundamentally alter the description. The main impact is a further stabilization of the impurity's spin by the exchange interaction of the local orbital electrons.\newline
Having tested our model in the way described above, we can now turn to studying the influence of a discretized host density of states on the spin polarization of the impurity. We will proceed in two steps. First, we will keep the host density of states quasi-continuous and introduce an energy gap at the Fermi level. Second the host density of states will additionally be discretized.
\begin{figure*}[th!]
\includegraphics[width=0.7\textwidth]{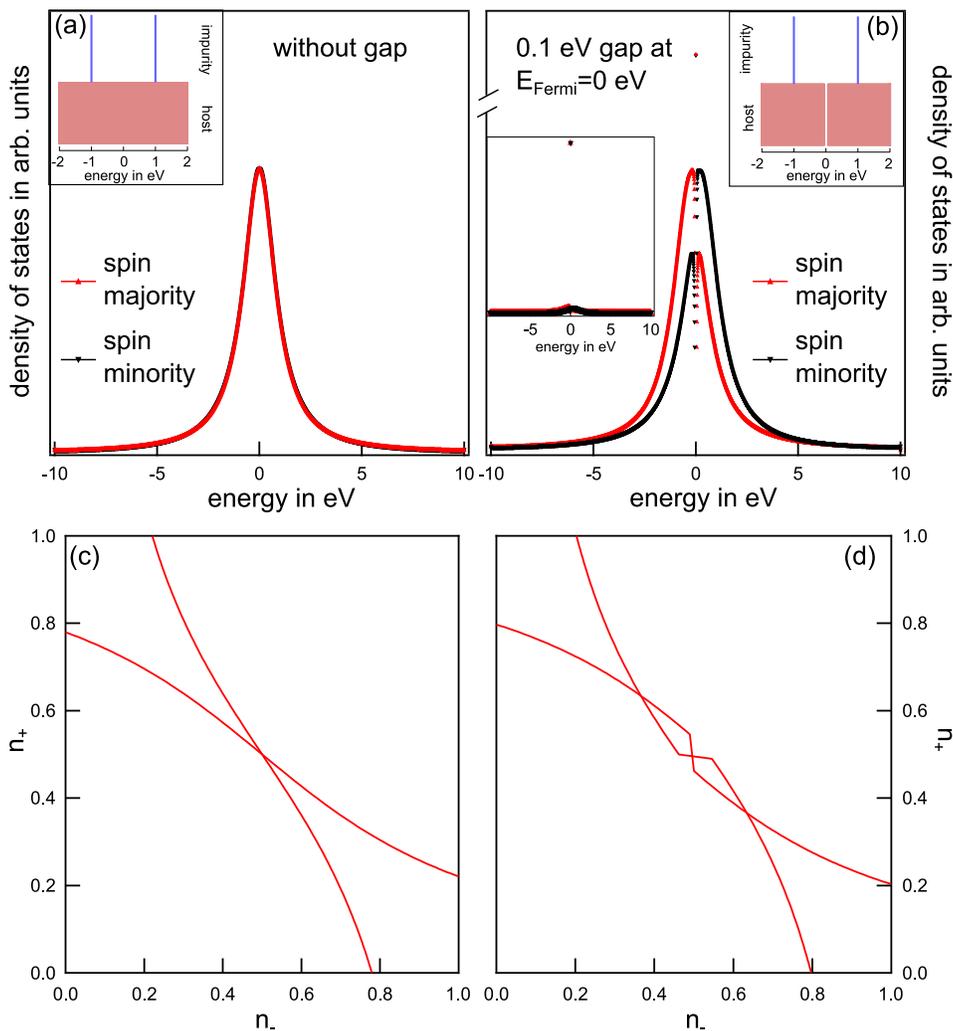}
\caption{\label{fig:VBS}Upper panels: Impurity state projected density of states obtained using the tight-binding model Hamiltonian, equation (\ref{eq:TB_Hamiltonian}), for an impurity interacting with a dense discrete host density of states ($\unit[180]{eV^{-1}}$) without a gap (a) and with a gap of $\unit[0.1]{eV}$ (b). The ordinate in panel (b) is interrupted, while the inset shows the impurity state projected density of states as calculated. Parameters $a=\unit[0.04]{eV}$ and $U_0=\unit[2]{eV}$ were kept constant. The insets show the uncoupled impurity and host density of states. Lower panels (c) and (d) show the resulting solution for the occupation numbers for spin-up and -down states. The impurity magnetization is restored when introducing a small gap in the host density of states, panel (d). The values used correspond to an Anderson criterion of $U_0/\Gamma=2.21<\pi$.}
\end{figure*}
\section{Energy Gap in the Host Density of States}
The influence of an energy gap in the host density of states on the total occupation of the impurity states has been studied in the seminal work of Haldane \cite{Haldane1976}, which is an extension of the Anderson impurity model. Haldane was able to explain the large variety of charge states that are observed in dilute magnetic semiconductors. However, the spin polarization was not addressed in Haldanes study.
\begin{figure}[ht!]
\includegraphics[width=0.9\columnwidth]{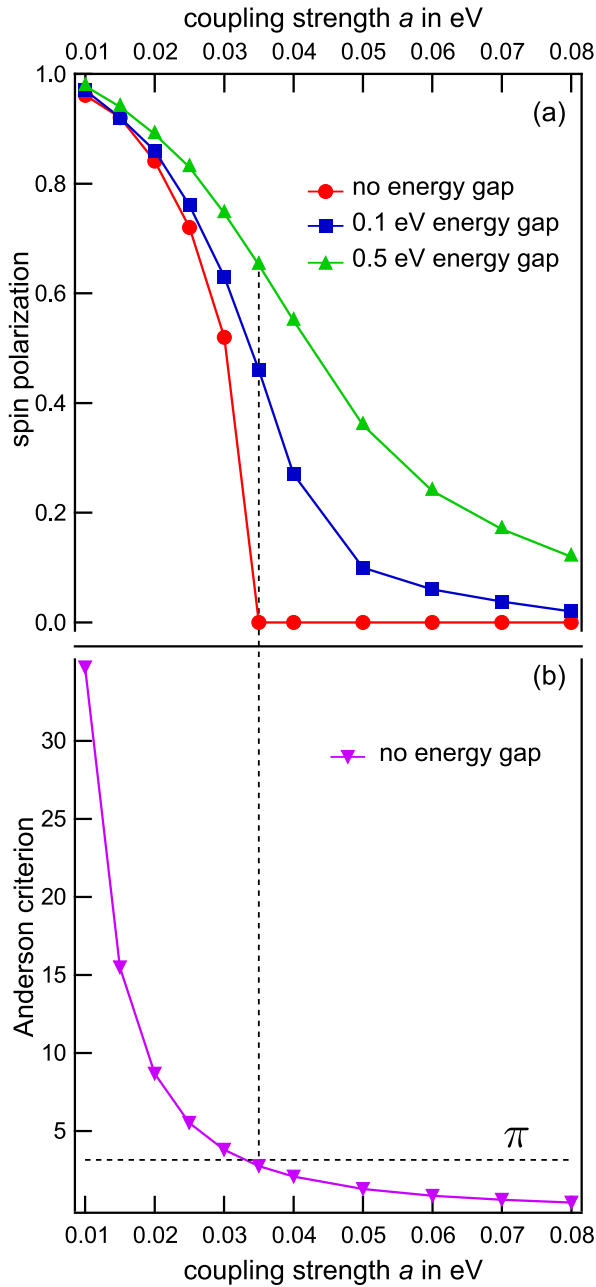}
\caption{\label{fig:Spin_a}Comparison of systems incorporating an energy gap (\unit[0.1]{eV} and \unit[0.5]{eV}) in the host density of states to a system exhibiting no energy gap. On-site Coulomb repulsion $U_0=\unit[2]{eV}$ and density of states $\unit[180]{eV^{-1}}$ were kept constant. Panel (a): Spin polarization as a function of coupling strength $a$. Only for very small coupling parameters similar spin polarizations can be found. The spin magnetic moment is quenched for $a \geq \unit[0.035]{eV}$ in the continuous band case, whereas it survives in presence of a gap. Panel (b): Anderson criterion drops below $\pi$ for coupling strengths $a \geq \unit[0.035]{eV}$ marking the magnetic-to-nonmagnetic transition.}
\end{figure}
\begin{figure}[ht!]
  \includegraphics[width=0.9\columnwidth]{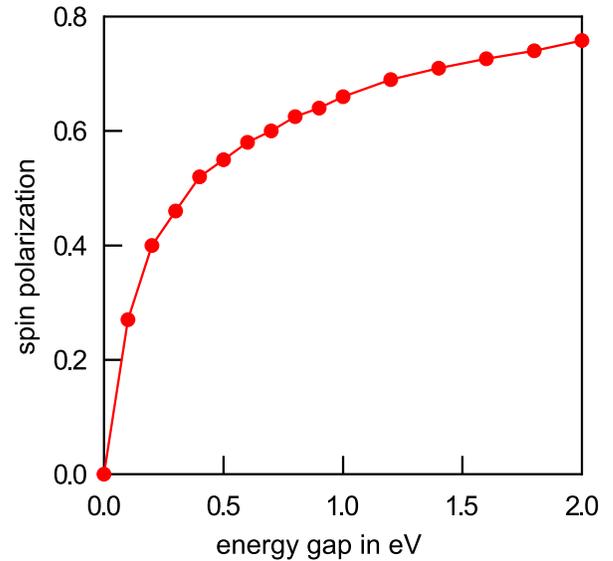}
  \caption{Spin polarization of the impurity as a function of the host energy gap at constant coupling parameter $a=\unit[0.04]{eV}$, on-site Coulomb repulsion $U_0=\unit[2]{eV}$ and host density of states of $\unit[180]{eV^{-1}}$.}
  \label{fig:Spin_Gap}
\end{figure}
In this study we will concentrate on the influence of a gap on the spin polarization of the system.\newline
Parameters of on-site Coulomb repulsion $U_0=\unit[2]{eV}$, host density of states of $\unit[180]{eV^{-1}}$ and coupling strength $a=\unit[0.04]{eV}$ were chosen so that the spin polarization of the system vanishes in the continuous band limit. The relative energies of the electronic states of the uncoupled host and impurity are depicted in the inset of Fig. \ref{fig:VBS} (a). Vanishing spin polarization of the system is indicated by the degeneracy of the virtual bound majority and minority spin states of the composite system, obtained from a self-consistent solution as described in the previous section and shown in Fig. \ref{fig:VBS} (a). Note that self-consistency is only reached for the trivial non-magnetic solution $n_+=n_-=0.5$, as can be seen in panel (c) of the same figure.\newline
However, upon introducing an energy gap in the host density of states at the Fermi energy as small as $\unit[0.1]{eV}$, \textit{cf.} inset of Fig. \ref{fig:VBS} (b), the spin polarization is restored. Both, majority and minority spin states no longer feature a lorentzian shape, but exhibit poles at the Fermi level as shown in Fig. \ref{fig:VBS} (b). This in turn leads to a transfer of density of states from the minority to the majority spin state resulting in a finite spin polarization. This can also be seen in Fig. \ref{fig:VBS} (d) where additionally to the non-magnetic solution $n_\pm=0.5$ magnetic solutions $n_+ \neq n_-$ can be found. The depicted curves are no longer in agreement with the analytical description within the Anderson impurity model. The deviation from the $\arctan$-function equation (\ref{eq:AIM_occ}) is most obvious in the regions exhibiting straight lines around $n_+=n_-=0.5$, which suppress the quenching of the local spin magnetic moment.\newline
These observations indicate that the simple criterion of the Anderson impurity model for the magnetic to non-magnetic transition, $U_0/ \Gamma=\pi$, does not hold if an energy gap is introduced in the host density of states. To be more specific, magnetic solutions can be found although the criterion $U_0/ \Gamma$ yields values smaller than $\pi$ in the continuous band limit. The stabilization of a magnetic solution by introduction of an energy gap to the host density of states seems to be a quite robust effect as can be seen from Fig. \ref{fig:Spin_a} (a). Here, a comparison of the spin polarization as a function of the coupling strength $a$ is shown between systems exhibiting an energy gap in the host density of states and a system lacking an energy gap. For very small coupling parameters $a$ the influence of the gap is negligible, as has already been shown experimentally \cite{Hirsch2013}. For larger coupling strengths $a$, however, a severe deviation between the system with and without energy gap can be observed, most strikingly at $a \geq \unit[0.035]{eV}$. For that particular set of parameters ($U_0=\unit[2]{eV}$, $a \geq \unit[0.035]{eV}$ and $\rho=\unit[180]{eV^{-1}}$) the spin polarization of the system without a gap in the host density of states vanishes while the spin polarization survives in case of the systems exhibiting a gap. The vanishing spin polarization can be associated to the drop of the Anderson criterion $U_0/\Gamma$ below $\pi$, marking the magnetic to non-magnetic transition in the Anderson impurity model as depicted in panel (b) of Fig. \ref{fig:Spin_a}.\newline
Although the magnitude of the spin polarization will depend on the particular choice of the parameters $U_0$ and $a$ as well as the host density of states, opening up an energy gap reliably introduces a spin polarization in the system. This is even true for comparable large coupling parameters $a$ corresponding to small values $U_0/\Gamma<\pi$ in the continuous band limit, \textit{cf.} panels (a) and (b) of Fig. \ref{fig:Spin_a}. Again, this clearly shows the breakdown of the simple criterion for the magnetic to non-magnetic transition as given in the Anderson impurity model.\newline
We can therefore state at this point that an energy gap in the host density of states has a profound influence on the spin polarization. We can quantify this influence by calculating the magnitude of the spin polarization as a function of the size of the gap. The spin polarization as a function of the energy gap is plotted in Fig. \ref{fig:Spin_Gap}. All other parameters are kept constant as before $U_0=\unit[2]{eV}$, $a=\unit[0.04]{eV}$, and density of states $\unit[180]{eV^{-1}}$. Again, opening up the gap immediately restores a spin polarization which then monotonically increases as a function of the gap in the host density of states. The dependence of the spin polarization on the coupling strength $a$ will be discussed in more detail in the next section.\newline
\section{Discrete Host Density of States}
\begin{figure}[t!]
\includegraphics[width=0.9\columnwidth]{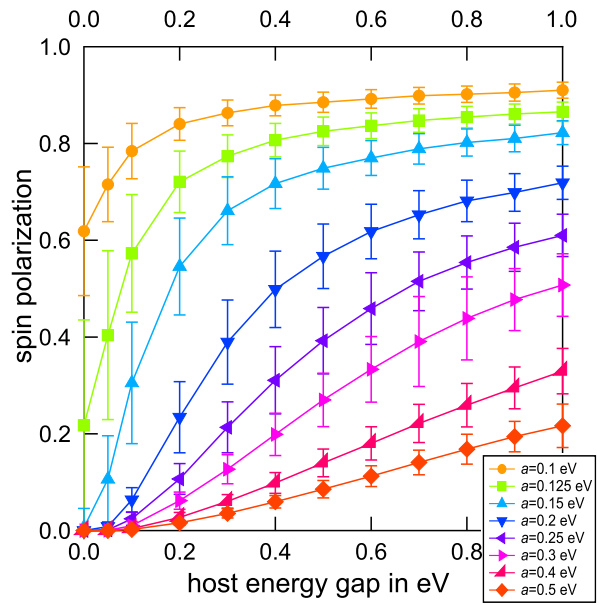}
\caption{\label{fig:Spin_vs_Gap}Spin polarization of a system consisting of 24 randomly distributed host states within \unit[20]{eV} as a function of the host energy gap for different coupling parameter $a$. Using an on-site Coulomb repulsion of $U_0=\unit[1]{eV}$.}
\end{figure}
The model Hamiltonian (\ref{eq:TB_Hamiltonian}) enables us to tackle not only bulk-like systems exhibiting a band gap but also to treat discrete energy levels of the host, which can be found, \textit{e.g.}, in isolated systems consisting of only a few atoms.\newline
It is reasonable to assume that the details of the spin polarization in such a system will depend on the exact relative arrangement of impurity and host electronic states. However, we will show here that the overall scaling of the spin polarization with the energy gap, which in case of finite systems is the HOMO-LUMO gap, of the host will hold also for a highly discretized host density of states.\newline
To this end we calculated the spin polarization of a system featuring 24 delocalized host states (12 occupied, 12 unoccupied) interacting with a single impurity state as a function of coupling strength $a$ as well as of the energy gap. The system introduced here could, \textit{e.g.}, be a $\mathrm{Na}_{12}$, $\mathrm{Au}_{12}$ or $\mathrm{Cu}_{12}$ host particle. As already pointed out the spin polarization will depend on the actual level arrangement. However, in order to get a detailed insight into the influence of the energy gap, for every set of parameters we generated thousand different host systems with randomly distributed host levels within a \unit[20]{eV} energy range \cite{Itoh2009} under the constraint to exhibit a certain energy gap at the Fermi level. The resulting density of states of $\unit[1.2]{eV^{-1}}$ compares well with the density of states at the Fermi level of a free electron gas, again neglecting level bunching due to electronic shell effects, in a coinage metal particle of size 12. The results of these calculations are depicted in Fig. \ref{fig:Spin_vs_Gap}, where the mean value of the system's spin polarization is plotted versus the energy gap. The standard deviation is given by the error bars. This was done for different coupling strengths $a=\unit[0.1-0.5]{eV}$ keeping $U_0=\unit[1]{eV}$ constant. The values for the coupling strength $a$ used in the calculation correspond to the parameters used in the previous section, since $a$ scales with the number of host states $N$ as $a=a_0/\sqrt{N}$.\newline
As can be seen from the figure, on average, the spin polarization scales inversely with the coupling strength $a$. But more importantly the spin polarization scales with the size of the energy gap, comparable to the behavior of a quasi-continuous band exhibiting a gap as presented in the previous section. The standard deviation is larger for small energy gaps and intermediate coupling strengths, \textit{cf.} Fig. \ref{fig:Spin_vs_Gap}, since for parameters close to the transition from a magnetic to a non-magnetic state the actual arrangement of the host and impurity states becomes naturally more important, in contrast to systems exhibiting very small or large coupling strengths $a$. In case of very weak interaction between host and impurity, hybridization is small independent of the relative arrangement of the levels. In the large coupling strength regime, hybridization is also mainly independent of the particular arrangement of the levels, since the host states are coupled to the impurity state disregarding their energetic separation.\newline
\section{Conclusion}
The influence of an energy gap in the host density of states as well as the discrete nature of a host density of states on the spin polarization of an impurity was studied using a tight-binding approach. The criterion for a transition from a magnetic to a non-magnetic system as stated within the Anderson impurity model is found not to be valid anymore. For cases where the magnetic moment of the impurity is quenched in a system having a continuous host density of states, we have shown that the opening of a gap can recover the spin magnetic moment. The size of the spin polarization scales with the size of the energy gap. This observation even holds for a discrete host density of states. On average the spin polarization follows the size of the energy (HOMO-LUMO) gap and the actual relative energetic position of impurity and host electronic states is of minor importance. This can severely influence the magnetic moment of impurities embedded in finite size matrices exhibiting a discretized density of states. Although their bulk counterpart may lack a magnetization, finite systems can exhibit a spin polarization, which can be tuned by the size of the host energy (HOMO-LUMO) gap. The described dependence is expected to be observable by studying, e.g. cobalt doped aluminum gas phase clusters combining x-ray magnetic circular dichroism \cite{Hirsch2009,Niemeyer2012,Zamudio-Bayer2013} and ultraviolet photoelectron spectroscopy \cite{Pettiette1988}.\\
Furthermore, the results even point at the possibility to switch the impurity's spin magnetic moment in a particular class of bulk host materials, i.e., materials exhibiting a Peierls transition. By passing through the Peierls transition temperature and thereby opening  and closing an energy gap at the Fermi level, respectively, the spin of an embedded impurity may be switched on and off.
\section{Acknowledgments}
KH thanks Linn Leppert for fruitful discussions.
\end{document}